\documentclass[lettersize,journal]{IEEEtran}
    \usepackage{booktabs}
    \usepackage{diagbox}
    \usepackage{makecell}
    \usepackage{array}
    \usepackage{graphicx,amssymb,amsmath}
    \usepackage{multicol}
    \usepackage[noadjust]{cite} 
    \usepackage{setspace}
    \usepackage{subfigure}
    \usepackage{graphicx}
    \usepackage{float}
    \usepackage {url}
    \usepackage{stfloats}
    \usepackage{amsthm,pifont}
    \usepackage{flushend}
    \usepackage{cases,subeqnarray}
    \usepackage{bm,multirow,bigstrut}
    \usepackage{amsmath, amsthm, amssymb}
    \usepackage{textcomp}
    \usepackage{latexsym,bm}
    \usepackage{booktabs}
    \usepackage{mathtools}
    \usepackage{dsfont}
    \usepackage{extarrows}
    \usepackage{epsfig}
    \usepackage{epsfig}
    \usepackage{epstopdf}
    \usepackage[table]{xcolor}
    \usepackage{colortbl} 
    \usepackage{multirow, hhline}
    \usepackage{xcolor}
    \usepackage{array} 
    \usepackage{algorithmicx,algorithm}
    \usepackage{hyperref}

    \theoremstyle{plain}

    \theoremstyle{plain}

    \usepackage{amsmath}

    \IEEEoverridecommandlockouts
\begin{document}
    %----------------------------title&author&thanks----------------------------
    % \title{Intelligence ISAC-enabled Metaverse:\\
    % From Human-avatar to object-digital twin}
    \title{Feature Engineering for Wireless Communications and Networking: Concepts, Methodologies, and Applications}
    
    \author{
    Jiacheng Wang, Changyuan Zhao, Zehui Xiong, Tao Xiang, Dusit Niyato,~\IEEEmembership{Fellow,~IEEE,} \\ Xianbin Wang,~\IEEEmembership{Fellow,~IEEE,} Shiwen Mao,~\IEEEmembership{Fellow,~IEEE,} Dong In Kim,~\IEEEmembership{Life Fellow,~IEEE,} 
    \vspace{-1cm}
    
    % <-this % stops a spaceU
    % \thanks{This work is supported in part by the National Natural Science Foundation of China (NSFC) under Grants No. 62102099, No. U22A2054, the Pearl River Talent Recruitment Program under Grant 2021QN02S643, and Guangzhou Basic Research Program under Grant 2023A04J1699; in part by the National Research Foundation, Singapore, and Infocomm Media Development Authority under its Future Communications Research \& Development Programme, DSO National Laboratories under the AI Singapore Programme (AISG Award No: AISG2-RP-2020-019 and FCP-ASTAR-TG-2022-003), and MOE Tier 1 (RG87/22)} 
    \thanks{J.~Wang, C.~Zhao, and D. Niyato are with the College of Computing and Data Science, Nanyang Technological University, Singapore (e-mail: jiacheng.wang@ntu.edu.sg, zhao0441@e.ntu.edu.sg, dniyato@ntu.edu.sg).}
    
    \thanks{T. Xiang is with College of Computer Science, Chongqing University, Chongqing 400044, China (e-mail: txiang@cqu.edu.cn).}
    
    \thanks{Z. Xiong is with the School of Electronics, Electrical Engineering and Computer Science (EEECS), Queen's University Belfast, Belfast, BT7 1NN, U.K. (z.xiong@qub.ac.uk).}

    \thanks{X. Wang is with the Department of Electrical and Computer Engineering, Western University, London, ON N6A 5B9, Canada (e-mail: xianbin.wang@uwo.ca)}
    
    \thanks{S. Mao is with the Department of Electrical and Computer Engineering, Auburn University, Auburn, USA (e-mail: smao@ieee.org).}
    
    \thanks{Dong In Kim is with the Department of Electrical and Computer Engineering, Sungkyunkwan University, Suwon 16419, South Korea (email: dongin@skku.edu).}
    %\thanks{M.~Zhou is with School of Communication and Information Engineering, Chongqing University of Posts and Telecommunications, Chongqing, China (e-mail: zhoumu@cqupt.edu.cn).}
    % \thanks{J. Kang is with the School of Automation, Guangdong University of Technology, Guangzhou, China (e-mail: kavinkang@gdut.edu.cn).}
    % \thanks{S. Cui is with the School of Science and Engineering (SSE) and the Future Network of Intelligence Institute (FNii), Chinese University of Hong Kong (Shenzhen), China (e-mail: shuguangcui@cuhk.edu.cn).}
    % \thanks{X. Shen is with the Department of Electrical and Computer Engineering, University of Waterloo, Canada (email: sshen@uwaterloo.ca).}
    % \thanks{P. Zhang is with the State Key Laboratory of Networking and Switching Technology, Beijing University of Posts and Telecommunications, China (e-mail: pzhang@bupt.edu.cn).}
    %\thanks{H. Vincent Poor is with the Department of Electrical and Computer Engineering, Princeton University, Princeton, NJ 08544 USA (e-mail: poor@princeton.edu).}
    }
\maketitle
    %----------------------------abstract----------------------------
    \begin{abstract}
     AI-enabled wireless communications have attracted tremendous research interest in recent years, particularly with the rise of novel paradigms such as low-altitude integrated sensing and communication (ISAC) networks. Within these systems, feature engineering plays a pivotal role by transforming raw wireless data into structured representations suitable for AI models. Hence, this paper offers a comprehensive investigation of feature engineering techniques in AI-driven wireless communications. Specifically, we begin with a detailed analysis of fundamental principles and methodologies of feature engineering. Next, we present its applications in wireless communication systems, with special emphasis on ISAC networks. Finally, we introduce a generative AI-based framework, which can reconstruct signal feature spectrum under malicious attacks in low-altitude ISAC networks. The case study shows that it can effectively reconstruct the signal spectrum, achieving an average structural similarity index improvement of 4\%, thereby supporting downstream sensing and communication applications.
        \end{abstract}
    %----------------------------keywords----------------------------
    %\newpage
    \IEEEpeerreviewmaketitle
    %---ction----------------------------
    \section{INTRODUCTION}

    In recent years, artificial intelligence (AI) has been increasingly integrated into various domains, in which feature engineering plays an indispensable role. Feature engineering refers to the process of extracting, transforming, and selecting informative variables from raw data, which becomes a foundational tool for different AI models across various fields. For example, generative diffusion models~\cite{wang2022guided} can encode images into a low-dimensional latent feature space, where the models learn feature patterns by focusing on the most important semantic attributes, thereby realizing realistic image generation. In discriminative tasks, such as image classification or speech recognition, well-designed features, such as scale-invariant features and histograms of gradient, can boost classification performance under diverse conditions.

    % Beyond computer vision, AI models have also rapidly penetrated into different wireless technologies, such as ISAC, from various perspectives. These AI-driven technologies play a vital role in modern communications, especially in emerging low-altitude scenarios. For example, in an ISAC system based on channel state information (CSI), researchers can extract various features from the raw CSI data acquired by UAVs, such as time-frequency spectrum, angle of arrival (AoA), and time of flight (ToF), and use these features to train AI models to complete tasks such as UAV localization and orientation recognition. In the network layer, the authors in~\cite{ye2024integrated} introduced a deep reinforcement learning-based ISAC framework designed for low-altitude economy scenarios, optimizing ground base station beamforming and UAV trajectories to maximize communication sum-rate while meeting sensing and operational constraints. Moreover, by modeling the mission as a Markov decision process, they further proposed a constrained noise-exploration policy and hierarchical experience replay to handle the episode-based structure effectively. 

    From a broader perspective, feature engineering plays a vital role in both generative AI (GenAI) and discriminative AI, which are widely used in wireless communications networks. For example, GANs have been trained to learn the distribution of measured channel responses and generate realistic synthetic channel samples for data augmentation~\cite{wei2022channel}. These methods typically first transform raw I/Q or channel measurements into feature spaces that capture essential channel characteristics, such as time–frequency spectrograms, enabling the generative model to effectively model and generate new wireless signals. In the discriminative paradigm, convolutional networks trained on spatial or spectral feature maps can pinpoint UAV or base-station locations from wireless data, and CNN-based classifiers achieve high accuracy in automatic modulation recognition when trained on 2D signal features, such as spectrogram images. In these cases, raw inputs are preprocessed into structured features, for example, by applying short-time Fourier or wavelet transforms. so that neural models can extract meaningful patterns. 

    The above analysis shows that feature engineering can make full use of domain-specific knowledge to obtain more appropriate features, thereby training a more efficient AI model for the task. Such approach brings the following significant advantages.

    \begin{itemize}
    \item \textbf{Better interpretability}: Engineered features used for training can embed physical insights (e.g., Doppler shifts, path delays) into the model. This alignment with wireless physics makes AI decisions easier to explain and verify.

    \item \textbf{Improved robustness}: Carefully selected features emphasize reliable signal traits (e.g., channel statistics, spectral envelopes). This helps models withstand noise, interference, and adversarial perturbations.

    \item \textbf{Efficient learning}: Compact, non-redundant features (e.g., principal components of CSI) lower input dimensionality and training data requirements. Models thus converge faster and demand fewer computational resource.

    \item \textbf{Performance boost}: Highlighting the most informative attributes, such as angular spreads, raises accuracy in tasks such as channel prediction, target detection, and spectrum reconstruction. Overall system reliability and throughput improve accordingly.
   
    \end{itemize}

    Based on the above benefits and motivations, this paper provides an in-depth and comprehensive investigation on the application of feature engineering in AI-driven wireless communications networks. Specifically, we first provide a comprehensive overview of feature engineering, including its fundamental concepts and key procedures. We then examine its applications in AI-driven wireless communications from the perspectives of resource optimization, signal processing, and network management, and further extend the discussion to its role in integrated sensing and communication (ISAC) networks. Building on this, we design a feature recovery framework based on a generative diffusion model. It takes the signal features extracted from the attacked ISAC network as input and reconstructs the attack-free signal features through two processes, including noise addition and guided de-noising. The performance of the proposed framework is finally analyzed through a case study to verify its effectiveness.

\section{Review of Feature Engineering}
\textit{Feature Engineering in AI and machine learning refers to the process of creating and refining input features from raw data to improve model performance.} Unlike learning features automatically given enough data, feature engineering injects domain knowledge into the pipeline, which is more useful when data is limited. This section presents a comprehensive overview of feature engineering including three main steps, including feature creation, transformation, and selection.
% \vspace{-0.5cm}
\subsection{Feature Creation}
Feature creation is the process of deriving informative attributes from raw data through mathematical transformations or filtering operations. The goal is to represent raw data in a new form that is more useful for AI models, typically by emphasizing important data characteristics while reducing its redundancy. For example, by replacing the Gaussian filter with a noise-suppressing filter, adding diagonal edge templates, and fine-tuning the threshold, the authors in~\cite{li2022improved} presented an improved Canny edge detection algorithm, which achieves more accurate and detailed edge detection in noisy images. In wireless communications, feature creation often involves converting time-domain signals into other domains to reveal hidden structure. For example, the authors in~\cite{wang2024acceleration} extracted a velocity-acceleration spectral feature from the CSI using a series of computations, including Fourier and keystone transform, and leveraged this feature to enable human fall detection in multi-target scenarios. Unlike deep learning approaches that automatically learn features, designed feature creation with domain knowledge offers greater specificity, providing more effective input support for downstream learning tasks.

\subsection{Feature Transformation}
After feature creation, feature transformation techniques are applied to modify or combine multiple features in ways that enhance their usefulness. One common transformation is normalization, which scales features to a comparable range so that no single feature dominates due to scale differences. For instance, the authors in~\cite{yang2023multi} compress CSI via normalization and projection transformation, which can enhance the orthogonality of CSI signal subspace and noise subspace, so that the following curve fitting can reach the optimal solutions faster. Another representative transformation is principal component analysis (PCA), which projects the features into a lower-dimensional space while preserving as much variance as possible. In~\cite{xia2021multiview}, the proposed multiview-PCA extends traditional PCA by segmenting tensors and applying orthogonal projections to preserve structure and maximize variation. By adjusting segmentation depth and direction, it generalizes both PCA-like and Tucker decomposition methods, offering improved adaptability for subspace recovery. In summary, feature transformation can further refine the feature set, through operations such as feature scaling, into a form that is more suitable for AI models.

\subsection{Feature Selection}
Feature selection is the process of identifying and selecting the most relevant subset of features from the overall feature pool. The key idea is that different features contribute differently to various tasks. Feature selection is realized by evaluating feature importance and pruning the less useful ones. One widely used technique is based on maximum relevance minimum redundancy (MRMR). For instance, the authors in~\cite{sun2021feature} first developed the margin-based fuzzy neighborhood radius, the fuzzy neighborhood similarity relationship, and the fuzzy neighborhood information granule. On this basis, they built the multi-label fuzzy neighborhood rough sets and used the MRMR model with label correlation to realize feature selection. MRMR efficiently balances relevance and redundancy, making it also well suited for high-dimensional wireless data such as channel measurements and sensor readings. Besides, wrapper methods are also commonly used. For instance, sequential forward selection starts with an empty feature set and iteratively adds the feature that yields the largest improvement in validation accuracy when combined with the already selected features. The outcome of feature selection is a more compact feature set that maintains and even improves the model’s predictive power.

    \begin{figure*}[htp]
    \centering
    \includegraphics[width=0.9\linewidth]{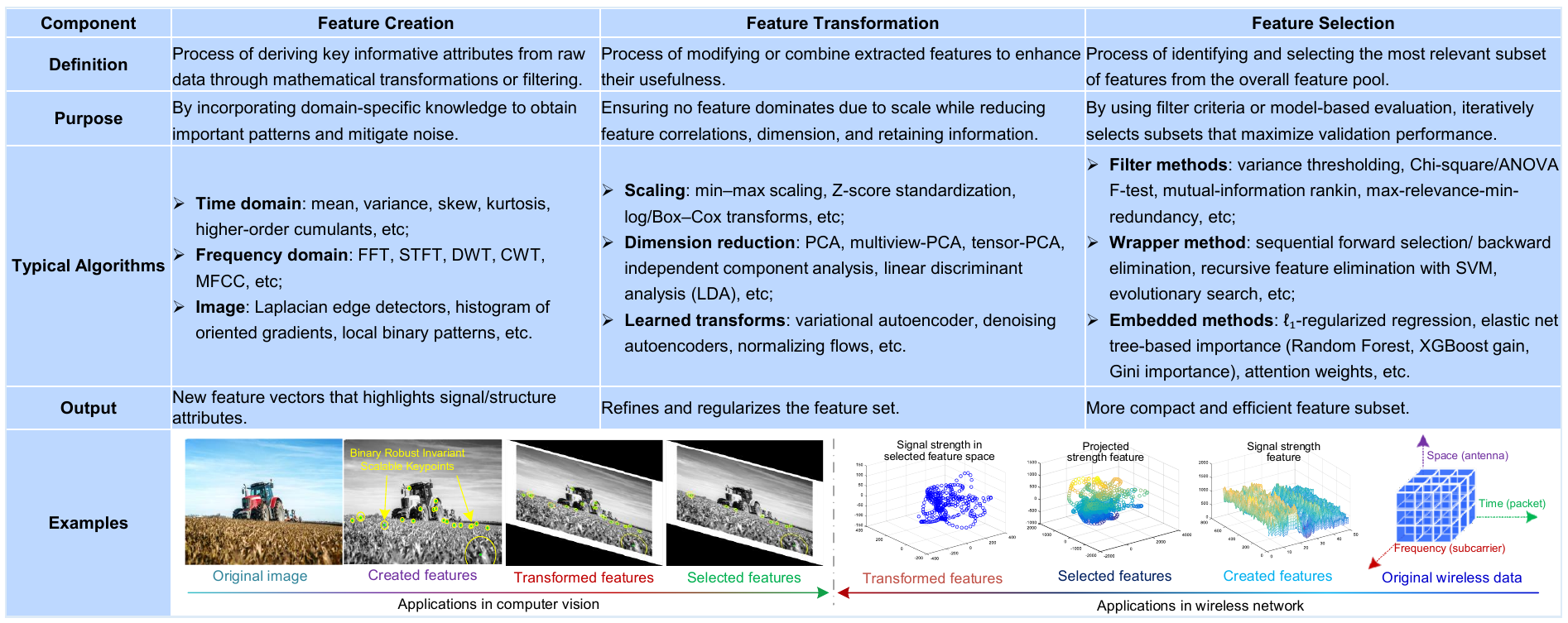}
    \caption{Summary of feature engineering, including the definition, purpose, typical algorithms, examples, etc. It is important to recognize that feature creation, transformation, and selection are not a rigid pipeline. Depending on the application, these operations may be freely combined, reordered, added, or omitted.}
    \label{FTE}
    \end{figure*}

\subsection{Lessons learned}
The way features are created, transformed, and selected has a direct impact on the performance of AI models. During these operations, some key points need attentions.

\begin{itemize}
\item With limited data, manually crafted features informed by domain expertise often yield better results than deep learning. With sufficient data, deep learning can discover more complex features but at higher cost. Moreover, features useful for one task may be irrelevant or misleading for another, so a few well-chosen features often work better than general ones.

\item The requirement and type of transformation often depend on the AI model. For example, models such as neural networks and SVMs perform better when input features have similar scales and distributions, making techniques including standardization or log transformation essential. Thus, it is important to align transformations with the model’s requirements.

\item Simple methods such as correlation ranking are fast but may miss important feature combinations, dropping those that are weak alone but strong together. In contrast, wrapper and embedded methods account for feature interactions and model specifics, often giving better results at the cost of more computation. Therefore, iterative and cross-validated approaches are preferred.

\item By modeling creation, transformation, and selection as an optimization problems, one can systematically solve them via methods such as $\ell_1$-regularization and mixed-integer programming. These methods provide a way to the searching and construction of high-quality feature sets.

\end{itemize}
In Fig.~\ref{FTE}, we briefly summarize feature engineering from different dimensions.

\section{Feature Engineering in AI-Driven Wireless Communications}
Feature engineering has extensive applications in AI-driven wireless communications. This section discusses several key areas in wireless networks where feature engineering is vital. 
% In ISAC networks, the data of interest, such as CSI, is high-dimensional, complex, and carries dual-purpose information. Hence, numerous techniques are developed to process CSI into useful features for tasks such as indoor localization, gesture recognition, human activity detection, and even vehicular sensing. In this section, we focus on feature extraction, augmentation, and classification to explore the use of feature engineering in ISAC systems.

\subsection{Signal Processing}
Applying AI models in signal processing for tasks, such as modulation recognition, requires distilling signals into effective feature sets. This often relies on hand-crafted features, such as higher-order cumulants, power spectral densities, and cyclostationary signatures of the received signal. For instance, the authors in~\cite{xing2024joint} present a framework to achieve complex signal detection and automatic modulation classification for multiple-signal coexisting environment. They input the original data stream as an I/Q sequence into the signal detection module, and the fast Fourier transform is used to obtain frequency-domain features. Here, a band prediction model is incorporated at the end of the detection module to generate predictions for the frequency location feature of the signal. Following this, a CNN-based module is utilized to predict the modulation pattern of each filtered sequence of pure signals, achieving impressive performance under a complex channel conditions. Similarly, for other physical layer issues, such as channel prediction, features such as the eigenvalues of the channel matrix and the delay spread extracted from impulse responses can better summarize the propagation environment, thereby improving the training and operation efficiency~\cite{castedo2002adaptive}. From these cases, one can see that while deep learning can learn features from raw signals, using domain knowledge to extract features based on specific communication scenarios is more efficient and results in solutions that are easier to deploy. 

\subsection{Resource Optimization}
Wireless networks must constantly manage limited resources across users and services. To handle this in complex and dynamic environments, deep reinforcement learning (DRL) has been widely adopted. In DRL-based optimization, a crucial step is constructing a meaningful representation of the network state for the learning agent. Feature engineering can encode network information, such as channel conditions, into a compact state vector for decision-making. For example, the authors in~\cite{zhang2020power} model a cognitive radio network, in which secondary users infer the primary user’s power allocation from sensor signal strengths and then adjust their own transmit power. They take the received signal strength as the state and introduce deep neural networks to solve the problem of infinite number of states. Here, the authors omitted the mobility, as it is less important in such scenario. Such strategic feature selection optimizes resource usage and can improve learning efficiency. By constructing these appropriate states, evaluations show that the agent can focus on the most influential metrics, leading to faster convergence. Hence, feature engineering can  identify the most relevant network metrics and combine them into a compact representation, enabling AI models to make near-optimal resource allocation decisions.

\subsection{Network Anomaly Detection}
Wireless networks generate vast amounts of operational data, and AI models can identify underlying patterns or issues by analyzing features derived from this data. Creating features that accurately reflect network states is therefore essential. For instance, the authors in~\cite{oleiwi2022mlts} propose an ensemble learning-based anomaly detection framework for communication networks. Specifically, raw data first undergoes normalization, duplicate removal, and categorical encoding, to ensure data quality and consistency. They then apply a hybrid method that combines correlation based feature selection with a random forest algorithm to identify a compact and highly relevant feature subset, including key network flow features such as ct\_srv\_src and ct\_dst\_src\_ltm, by evaluating interfeature redundancy and class relevance. Finally, modified ensemble classifiers, including random forest and support vector machine (SVM) with adaboosting and bagging, are used with a voting strategy for intrusion detection. The evaluation demonstrates that such method can achieve up to 99.6\% accuracy across the NSL\_KDD, UNSW\_NB2015, and CIC\_IDS2017 datasets. Besides that, metrics including user mobility patterns and cell load can also be combined into features that AI models use to anticipate and mitigate issues, such as the cell outage. 

In summary, feature engineering plays a vital role in enabling AI across wireless networks, spanning from the physical layer to network optimization and management. Although techniques may differ, the core idea remains the same by focusing on extracting the most useful features for the AI model. This approach enhances system performance while reducing the resource burden on wireless networks.

\section{Feature Engineering in ISAC }
\subsection{Applications in ISAC}
ISAC can be applied to various scenarios, and the tasks corresponding to each scenario require different domain-specific feature representations to highlight relevant signal patterns

In indoor ISAC, fine-grained gesture and human activity recognition typically rely on time-frequency features extracted through algorithms such as wavelet transforms, which inherently capture the micro-Doppler frequency shifts caused by moving body parts. Moreover, researchers proposed more advanced representations, such as the body-coordinate velocity profile that combines Doppler with orientation and location information, which can provide more distinctive patterns. Hence, AI models trained on these features can effectively recognize various gestures and activities, achieving impressive accuracy. In multi-person and broader scenarios, additional spatial features, such as range and AoA, become more important. These features can be extracted via various array signal processing techniques, such as spatial filtering and parameter estimation. As an example, the authors in~\cite{wang2024generative} captured CSI with a uniform linear array and combined the MUSIC algorithm with diffusion models to estimate the time-of-flight and AoA of signals under conditions where antenna spacing exceeds half a wavelength. Subsequently, by clustering these estimated parameters, they realized effective human flow detection. Evaluations based on practical downlink communication signals demonstrate that this method achieves a sub-stream size detection accuracy of up to 91\%.

In low-altitude air space, the DRL is often used for the UAV-assisted ISAC networks. In~\cite{zhu2024resource}, a DRL framework is proposed for multi-UAV resource allocation, where each UAV simultaneously serves as a communication node and radar. Here, the designed feature state includes each UAV's 2D/3D location and velocity, its residual energy level, the locations of sensing targets, the current resource allocation strategy, and even the previous reward. By encoding UAV's position and velocity alongside communication-radar resource status, the agent’s state representation preserves critical features that influence both wireless channel quality and sensing performance. Such comprehensive representation enables the agent’s reward to account for both communication throughput and sensing accuracy, allowing the DRL policy to balance target detection and communication under dynamic conditions. Similarly, the authors in~\cite{liu2025integrated} employed a DRL approach to jointly optimize a UAV’s trajectory and ISAC signaling, where the feature state vector includes the UAV’s 3D coordinates, estimated CSI to ground users and reflecting surfaces, and distances to targets. Such informative features allow the agent to infer crucial context, including the distance between UAV and warden, the UAV-to-base link geometry, and the energy left for maneuvering or data transmission. This enables system to achieve more robust decision-making and improved trade-offs between communication and sensing accuracy in real time.

\begin{figure*}[htp]
    \centering
    \includegraphics[width=0.9\linewidth]{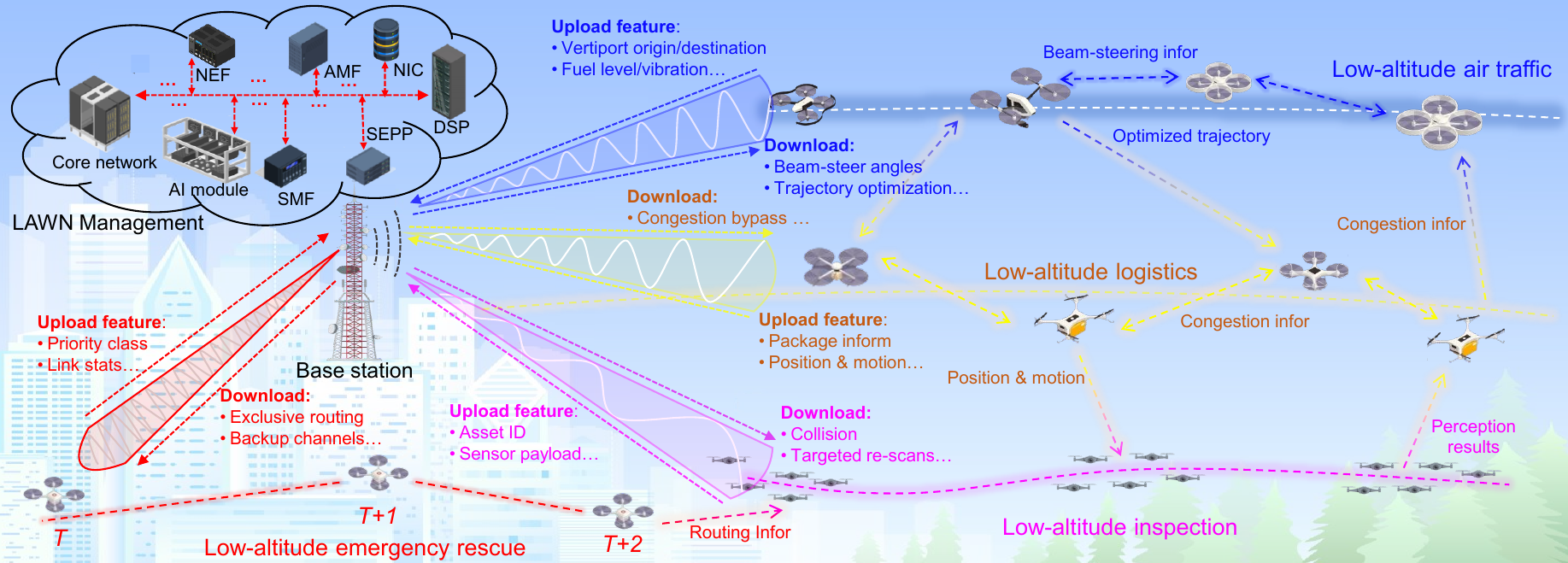}
    \caption{Feature engineering in low-altitude ISAC networks. In such scenario, aircraft for different tasks upload various feature data to the server. The server processes and analyzes these features and then makes decisions based on analysis result, and other factors such as current network conditions, air-traffic status, and so forth. Finally, server returns the results to the aircraft. Digital signal processing (DSP), security edge protection proxy (SEPP), network intelligence controller (NIC), access and mobility function (AMF), session management function (SMF), and network exposure function (NEF). }
    \vspace{-0.2cm}
    \label{APL}
    \end{figure*}

These examples illustrate that for various tasks, researchers can design efficient techniques for feature creation, transformation, and selection tailored to specific requirements, thereby providing concise inputs to AI models. Fig.~\ref{APL} shows some use cases of feature engineering in low-altitude ISAC scenarios. Nevertheless, in practice, the application of such feature engineering still faces significant challenges.

\subsection{Challenges}
Despite the benefits of feature engineering for low-altitude ISAC, its effectiveness is limited by fundamental practical constraints. Two of the most critical issues are shown below. 

\begin{itemize}

    \item \textbf{Vulnerability to malicious attacks.} One major challenge is the vulnerability of low-altitude ISAC networks to malicious attack in open air space. The open and shared nature of wireless channels in low-altitudes airspace makes ISAC networks susceptible to jamming or spoofing attacks by adversaries. Such malicious attacks can severely contaminate the wireless signals, distorting or masking the underlying features that the network tries to extract. For such issue, one of the solutions is to apply appropriate feature transformations, such as feature recovery that filter out or compensate for interference, thereby reconstructing feature representations.

    \item \textbf{High dynamics.} Low-altitude operations face constant shifts in terrain, obstacles, and channel quality, hence AI models must update quickly to stay effective~\cite{yuan2025ground}. Targeted feature engineering, especially careful selection and prioritization, keeps the state space clean by emphasizing stable, high-value descriptors and omitting redundant ones. A streamlined feature set cuts computation, speeds online learning, and preserves sensing and communication accuracy as conditions evolve. In turn, the ISAC system can adapt in real time and maintain robust performance amid continual environmental change.
    
\end{itemize}

Addressing these challenges is crucial to ensuring that low-altitude ISAC networks remain effective and reliable.

\section{Feature Recovery Under Malicious Attacks}
As discussed above, robust features engineering technologies are critical for low-altitude ISAC systems. Hence, in this section we propose a novel feature transformation technology, which can recover the feature spectrum under malicious attacks, thereby supporting downstream ISAC applications.
\vspace{-0.2cm}
\subsection{The Proposed Framework}
\begin{figure*}[htp]
    \centering
    \includegraphics[width=0.9\linewidth]{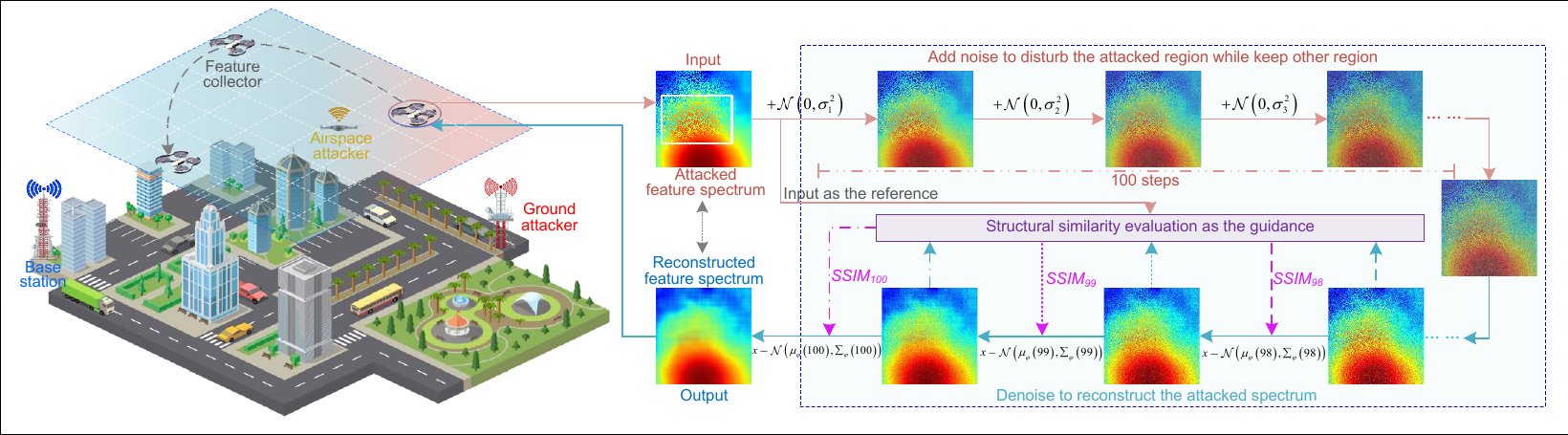}
    \caption{The proposed framework. It includes two transformation processes, i.e., the diffusion and denoising process. In the diffusion process, it adds noise to transform the input attacked signal feature spectrum to noisy spectrum. Then, in the denoising process, the injected noise and the interference introduced by the attack are eliminated via denoising, thereby reconstructing the clear feature spectrum.}
    \label{AIGX-PHY6}
    \vspace{-0.3cm}
    \end{figure*}

The proposed received-signal-strength indicator (RSSI) feature recovery framework is based on the generative diffusion model that contains two Markov chain-based transformation processes, as shown in the Fig.~\ref{AIGX-PHY6}. The forward process corrupts the RSSI spectrum by adding Gaussian noise, transforming the input into the noisy signal feature. The reverse process attempts to reverse this transformation, remove the noise and generate clear spectrum. More concretely, when the low-altitude ISAC network is under malicious attacks, the framework first feeds the extracted signal RSSI spectrum into the forward process, where the controlled Gaussian noise is injected to submerge the perturbations caused by attacks. This is essentially the feature creation process, which synthesize a richer and multi-scale neighborhood around the input RSSI spectrum via noise injection, increasing the opportunity that later reverse process can find a path back to the clean spectrum manifold. Subsequently, the reverse process iteratively denoise the signal RSSI spectrum, eliminating both the injected noise and the perturbations caused by attack to reconstruct clean signal RSSI spectrum. During this process, the framework employs the U-Net with the attention mechanism for feature extraction while integrating time embeddings. Here, the attention mechanism sharpens the network’s focus on details of the RSSI spectrum, the U-Net enhances multi-scale feature extraction capability, while the time embeddings ensure that the learned representations remain temporally consistent with the input RSSI spectrum. After that, the trained U-Net, combined with the denoising procedure, serves as a feature transformation module that decodes latent variables into the target domain, thereby reconstructing a clean signal spectrum. 

Here, a critical parameter is the number of diffusion steps in the forward process. Concretely, too many steps submerge not only the perturbations but also the underlying signal patterns, while too few steps may fail to submerge the perturbations. To balance these effects, our framework adopts two strategies. First, it can perform multiple rounds of forward-reverse process, each with a moderate number of steps. This is more reliable, as each round eliminates only a fraction of the perturbations. Second, it employs the signal features with perturbations as guidance during the reverse process, allowing the patterns of the input signal spectrum to be preserved.

% These two strategies offer the following advantages.

    % \begin{itemize}
    % \item  This can eliminate attack-induced perturbations through several rounds of forward-reverse transformation. Compared with a single round of transformation with a large number of steps, this strategy is more reliable, as each round eliminates only a fraction of the perturbations. 
    
    % \item By using the signal features with perturbations as the guidance in the denoising process, it can eliminate the perturbations while maintaining consistency with the pattern of the input signal features, ensuring the validity of the reconstructed signal features.
    % \end{itemize}
    
In the above described process, the diffusion model is pretrained based on clean signal features, hence the reconstructed feature is free of perturbations while preserving consistent underlying signal feature patterns.
% \vspace{-0.5cm}

% Throughout the process, the proposed strategy balances and coordinates the these modules by optimizing resource allocation, with the goal of minimizing the distribution gap in the feature space between the augmented data and the test data. This ensures that the augmented data can effectively support downstream tasks in ISAC networks. The proposed framework offers the following advantages.

% \begin{itemize}
% \item The proposed framework considers both the quantity and quality of feature samples during the augmentation process, resulting in more balanced and comprehensive data augmentation.

% \item Through the proposed optimization strategy, limited computational resources can be more effectively allocated across different modules, thereby achieving more efficient utilization.

% \item The proposed framework ensures that the augmented data remains consistent with the test data within the feature domain, which guarantees the effective performance of downstream tasks.
% \end{itemize}

% Note that the proposed framework can integrate multiple feature creation algorithms to support various communication and sensing tasks. In such cases, the framework selects suitable algorithms based on available resources and the downstream tasks, ensuring effective data augmentation while minimizing the distribution gap between augmented and test data.

\vspace{-0.2cm}
\subsection{Case Study}
Consider an open low-altitude airspce, in which an eVTOL samples the received-signal-strength indicator (RSSI) transmitted by ground base stations and constructs the RSSI feature spectrum. These spectra can be used to support numerous low-altitude ISAC functions, including localization, flight-trajectory optimization, and communication-quality assessment. The eVTOL is assumed to hover 100 m above ground, collecting RSSI over a region discretized into a 128 × 128 grid, with each cell measuring 4 m × 4 m. The base station located on the ground transmits the signal at 1.8 GHz with the power of 20 dBm. Here, both line-of-sight (LoS) and non-line-of-sight (NLoS) links are considered. The probability of the LoS path, denoted as $P_{LoS}$, can be computed using the 3GPP TR 38.901 channel model where the fixed coefficients are 9.61 and 0.16, while $P_{NLoS}=1-P_{LoS}$. Beside, the path loss follows a logarithmic attenuation model and the LoS and NLOS path-loss exponents are 2.2 and 3.8, respectively. For NLoS propagation, shadow fading is further modeled by a zero-mean Gaussian random variable with a variance of 6 dB and a de-correlation distance of 50 m. 

Building upon this, we assume the eVTOL experiences attacks during signal acquisition, with attack probabilities set at 0.3, 0.4, 0.5, 0.6, and 0.7. Two attack modes are considered. 
\begin{itemize}
\item Ground-based attack: an attacker on the ground transmits the interfering signal with power of 10 dBm.

\item Airborne attack: An attacker at an altitude of 100 m transmits an interfering signal with a power of 10 dBm, maintaining a 50 m distance from the target.
\end{itemize}
After that, the corrupted spectra collected by the eVTOL are fed into the proposed method, which eliminates the perturbations to reconstruct the original feature spectrum.
\vspace{-0.2cm}
\subsection{Performance Evaluation}
\begin{figure*}[htp]
    \centering
    \includegraphics[width=0.95\linewidth]{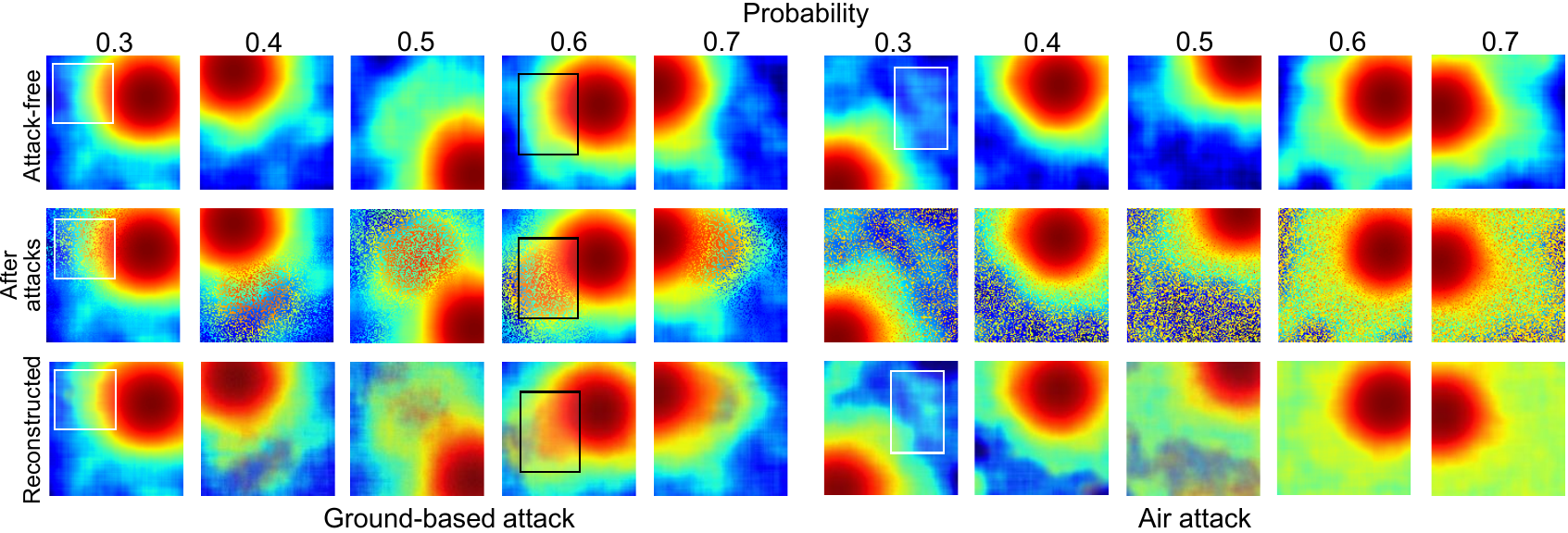}
    \caption{Signal feature spectrum reconstruction results under different conditions.}
    \label{SPC}
    \end{figure*}

Experimental results in Fig.~\ref{SPC} clearly show that when the attack probability is below 0.5, the proposed framework can effectively reconstruct the feature spectrum and recover local features in the attacked regions, as highlighted by the white boxes. Once the attack probability exceeds 50\%, performance declines. In particular, the method struggles to accurately recover details in areas where the interfering signal has higher power, as indicated by the black boxes. Comparing the two attack methods further reveals that feature spectrum is more challenging when the attacker is in the air. This is because the air attacker is within LoS and close to the eVTOL, causing stronger and more spatially extensive interference signal reaching the eVTOL, which causes more seriously perturbations.

\begin{figure}[t]
    \centering
    \includegraphics[width=0.4\textwidth]{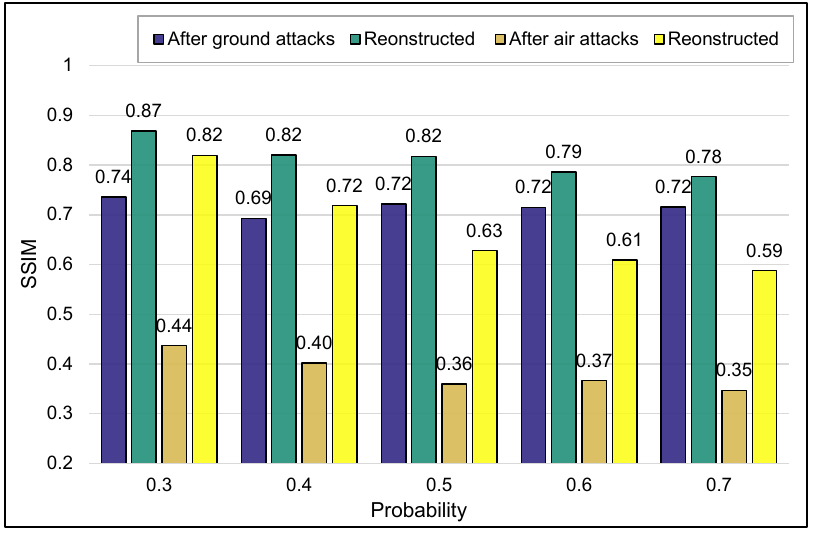}%scale=0.5
    \caption{SSIM between the attack-free feature spectra and reconstructed spectra under different cases.}
    \label{SSIM}
    \end{figure}

Using the attack-free feature spectrum as the reference, we computed the structural similarity index (SSIM) of the reconstructed spectra under various attack scenarios, and the results are shown in Fig.~\ref{SSIM}. The experimental curves show a marked SSIM degradation for spectra subject to attacks. The decline is most pronounced when the attacker operates at low-altitude air space, where the SSIM of the RSS spectrum falls to as low as 0.35, indicating severe disruption of its structural characteristics and posing substantial risks to downstream ISAC applications. After applying the proposed reconstruction method, the SSIM of the feature spectrum increases across all scenarios. Specifically, as shown in Fig.~\ref{SSIM}, the smallest improvement is 8\% (ground-based attack with probability of 0.3), the largest improvement reaches 87\% (airborne attacker with probability of 0.3), and the average improvement is 44\%. These results confirm that the proposed feature transformation framework can effectively reconstruct the signal feature spectrum that closely matches the attack-free baseline, validating its effectiveness.

    \vspace{-0.2cm}
  \section{Future Directions}
    \subsection{Feature Security in Low-altitude ISAC} 
    Future research should harden low-altitude ISAC against intentional interference that corrupts feature vectors. One strategy is to use adversarial-training or certified-robust learning, so that small feature perturbations fail to mislead sensing or communication modules. Besides that, multi-agent cross-validation, where UAVs and ground nodes compare captured features independently and then expose inconsistent measurements, thereby triggering rapid mitigation. Together, these directions can turn feature security from a reactive patch into a built-in system capability.
    
    \subsection{Feature Enhancement for Low-altitude ISAC}
    Improving feature quality is essential for reliable ISAC in dynamic airspaces. Generative models such as diffusion or GAN architectures can synthesize additional training samples under varying altitudes, clutter levels, and weather, reducing over-fitting to sparse real data. Beside, cross-modal feature fusion, such as combining wireless signal and vision, may uncover complementary features that boost both sensing resolution and communication robustness. In such case, feature-selection mechanisms must prioritize the most informative subsets when computational resources are limited. These enhancement strategies can boost model accuracy without demanding costly hardware upgrades.

    \subsection{Resource Allocation for Low-altitude ISAC}
    Resource management must balance sensing accuracy with communication throughput under tight energy and spectrum budgets. Learning-based schedulers, such as multi-agent DRL, can adapt allocations on-the-fly to channel fades, target maneuvers, or unexpected jamming. To ensure the effectiveness of the above methods, corresponding techniques need to be designed to obtain concise and efficient features. Such technology needs to be adaptable in order to dynamically adjust feature extraction strategies based on changing low-altitude environments. Meanwhile, the extracted features need to accurately describe the sensing and communication performance under different external conditions and resource states, ensuring fair and reliable resource allocation.
    
    \section{Conclusion}
    This paper offers a comprehensive overview of feature engineering and examines its role in AI-based wireless communications, with particular emphasis on low-altitude ISAC networks. Our investigation shows that feature engineering permeates the entire AI-based communication system, providing strong support for tasks ranging from signal processing to network-level resource optimization and management. On this basis, we have proposed a GenAI-based framework, which can restore signal RSS features spectrum when the low-altitude ISAC networks is attacked, providing support for downstream ISAC applications. The evaluations confirmed the framework’s effectiveness and further highlight the importance of feature engineering techniques in AI-based communication systems.
    
    % \newpage
     \bibliographystyle{IEEEtran}
     \bibliography{IEEEabrv,Ref}

    \end{document}